\documentclass[]{aa}
\usepackage{graphics}
\usepackage{rotating}
\usepackage{natbib}
\bibpunct{(}{)}{;}{a}{}{,} 
\newcommand{\ratio} {N({\rm H}_2) / I_{\rm CO}}
\newcommand{\kms}   {{\rm \, km \, s^{-1}}}
\def\ga{\lower.5ex\hbox{$\; \buildrel > \over \sim \;$}}
\def\la{\lower.5ex\hbox{$\; \buildrel < \over \sim \;$}}
\begin{document}
%
%
%


\def\jnl@style{\it}
\def\ref@jnl#1{{\jnl@style#1}}

\def\aj{\ref@jnl{AJ}}                   
\def\araa{\ref@jnl{ARA\&A}}             
\def\apj{\ref@jnl{ApJ}}                 
\def\apjl{\ref@jnl{ApJ}}                
\def\apjs{\ref@jnl{ApJS}}               
\def\ao{\ref@jnl{Appl.~Opt.}}           
\def\apss{\ref@jnl{Ap\&SS}}             
\def\aap{\ref@jnl{A\&A}}                
\def\aapr{\ref@jnl{A\&A~Rev.}}          
\def\aaps{\ref@jnl{A\&AS}}              
\def\azh{\ref@jnl{AZh}}                 
\def\baas{\ref@jnl{BAAS}}               
\def\jrasc{\ref@jnl{JRASC}}             
\def\memras{\ref@jnl{MmRAS}}            
\def\mnras{\ref@jnl{MNRAS}}             
\def\pra{\ref@jnl{Phys.~Rev.~A}}        
\def\prb{\ref@jnl{Phys.~Rev.~B}}        
\def\prc{\ref@jnl{Phys.~Rev.~C}}        
\def\prd{\ref@jnl{Phys.~Rev.~D}}        
\def\pre{\ref@jnl{Phys.~Rev.~E}}        
\def\prl{\ref@jnl{Phys.~Rev.~Lett.}}    
\def\pasp{\ref@jnl{PASP}}               
\def\pasj{\ref@jnl{PASJ}}               
\def\qjras{\ref@jnl{QJRAS}}             
\def\skytel{\ref@jnl{S\&T}}             
\def\solphys{\ref@jnl{Sol.~Phys.}}      
\def\sovast{\ref@jnl{Soviet~Ast.}}      
\def\ssr{\ref@jnl{Space~Sci.~Rev.}}     
\def\zap{\ref@jnl{ZAp}}                 
\def\nat{\ref@jnl{Nature}}              
\def\iaucirc{\ref@jnl{IAU~Circ.}}       
\def\aplett{\ref@jnl{Astrophys.~Lett.}} 
\def\apspr{\ref@jnl{Astrophys.~Space~Phys.~Res.}}
\def\bain{\ref@jnl{Bull.~Astron.~Inst.~Netherlands}} 
\def\fcp{\ref@jnl{Fund.~Cosmic~Phys.}}  
\def\gca{\ref@jnl{Geochim.~Cosmochim.~Acta}}   
\def\grl{\ref@jnl{Geophys.~Res.~Lett.}} 
\def\jcp{\ref@jnl{J.~Chem.~Phys.}}      
\def\jgr{\ref@jnl{J.~Geophys.~Res.}}    
\def\jqsrt{\ref@jnl{J.~Quant.~Spec.~Radiat.~Transf.}}
\def\memsai{\ref@jnl{Mem.~Soc.~Astron.~Italiana}}
\def\nphysa{\ref@jnl{Nucl.~Phys.~A}}   
\def\physrep{\ref@jnl{Phys.~Rep.}}   
\def\physscr{\ref@jnl{Phys.~Scr}}   
\def\planss{\ref@jnl{Planet.~Space~Sci.}}   
\def\procspie{\ref@jnl{Proc.~SPIE}}   

\let\astap=\aap
\let\apjlett=\apjl
\let\apjsupp=\apjs
\let\applopt=\ao

\title{Colliding molecular clouds in head-on galaxy collisions}
\author{J. Braine\inst{1}, U. Lisenfeld\inst{2}, P.-A. Duc\inst{3}, E. Brinks\inst{4},
V. Charmandaris\inst{5,6}, S. Leon\inst{2}}
\offprints{Jonathan~Braine, braine@obs.u-bordeaux1.fr}
\institute{ Observatoire de Bordeaux, UMR 5804, CNRS/INSU, B.P.
89,  F-33270 Floirac, France
\and Instituto de Astrof\'{\i}sica de Andaluc\'{\i}a, CSIC, Apdo. Correos 3004,
18080 Granada, Spain
\and CNRS URA 2052 and CEA/DSM/DAPNIA, Service d'Astrophysique, Saclay,
91191 Gif sur Yvette cedex, France
\and INAOE, Apdo. Postal 51 y 216, Puebla, Pue. 72000, Mexico
\and Cornell University, Astronomy Department, Ithaca, NY 14853, USA
\and Chercheur Associ\'e, Observatoire de Paris, LERMA, 61 Av. de l'Observatoire,
75014 Paris, France}
\date{Received / Accepted}
\authorrunning{Braine et al.}
\titlerunning{}
\abstract{We present further observations of molecular gas in
head-on collisions of spiral galaxies, this time of the
CO($J=1\rightarrow 0$) and
CO($J=2\rightarrow 1$) lines in the UGC~813 -- UGC~816 system.
UGC 813/6 are only the second known example of head-on spiral-spiral
collisions, the first example being the UGC 12914/5 pair.
Strong CO emission is present in the bridge between UGC 813 and 816,
unassociated with stellar emission, just as in UGC 12914/5.
The CO emission from the UGC 813/6 bridge, not counting the emission from
the galaxies themselves, is at least that of the entire Milky Way.
Collisions of gas-rich
spirals are really collisions between the interstellar media (ISMs)
of the galaxies.  We show that collisions between molecular clouds
bring H$_2$ into the bridge region.  Although the
dense clouds are ionized by the collisions, they cool and recombine very
quickly and become molecular again even before the galactic disks separate.
Because the clouds acquire an intermediate velocity post-collision, they
are left in the bridge between the separating galaxies.  The star
formation efficiency appears low in the molecular clouds in the bridges.
We speculate that the pre-stellar cores in the molecular clouds may
expand during the cloud collisions, thus retarding future star formation.
Because the ISM-ISM collisions discussed here require a very small impact
parameter, they are rare among field spirals.  In clusters, however,
these collisions should be an important means of ejecting enriched gas
from the inner parts of spirals.
\\
}

\maketitle

\section{Introduction}

A new class of colliding galaxies has recently come to be recognized --
galaxy encounters in which the neutral interstellar media (ISMs) of the
systems hit each other in collisions with very small impact parameters,
irrespective of the relative inclinations of the galactic planes.
These systems are characterized by a bridge with strong synchrotron emission
and abundant atomic hydrogen \citep[][ hereafter C93]{Condon93}.
Two galaxy pairs have clearly
undergone such a collision -- the UGC 12914/5 \citep[C93,][]{Jarrett99}
and UGC 813/6 \citep[][ also VV 769]{Condon02} pairs.  Contrary to initial
expectations (C93), extremely strong CO emission was detected throughout the
UGC 12914/5 bridge \citep{Braine03,Gao03}, showing that a large quantity of
molecular gas is present as well.  In fact, the CO emission from the
UGC 12914/5 bridge was found to be nearly 5 times that of the entire Milky
Way.  Due to the appearance of the synchrotron brightness contours and
the magnetic field, the
UGC 12914/5 system is also known as the "Taffy galaxies".  The UGC 813/6
pair shares these properties and thus is a member of the class of taffy
galaxies.  In this work, we show that the UGC 813/6 bridge contains
a lot of molecular gas, again like UGC 12914/5.  The centers of UGC 813
and 816 were previously observed in CO(1--0) by \citet{Zhu99}.

It is commonly believed that molecular clouds are too small and dense
to hit each other \citep[{\it i.e.} they have a low filling factor,
{\it e.g.} ][]{Jog92} or
to be affected by collisions with more diffuse atomic or ionized
hydrogen clouds.  In addition to presenting the new results for UGC 813/6,
the goal of this work is to show that indeed collisions of molecular
clouds are plausible and even inevitable in collisions of gas-rich galaxies.
As the surface filling factor of the molecular gas in the spiral disks
increases, a head-on collision of the galaxies becomes radically more efficient
at drawing dense gas out of the parent disks and into the region between
the two galaxies which separate after collision.  The fate of the gas
is unknown and probably quite dependent on whether the galaxies merge
post-collision or not.  Such collisions are, however, likely the only
means of forcing much of the gas out of the inner regions of spirals
because tidal forces primarily affect the less bound external regions.

The stars are not affected by collisions of the ISMs of the galaxies
and of course stars do not collide with each other.  However, the optical
appearance can be perturbed by the tidal forces which affect both stars and
gas.  The slower the relative velocities of the galaxies in the collision,
the stronger the effect of the tidal forces will be.

The UGC 813/6 bridge system is actually part of a triplet, the third
member being CGCG 551-011 about 50 kpc away and at the systemic velocity
of the UGC 813/6 pair, about 5200 km s$^{-1}$.  Following \citet{Condon02}
we use the optical convention ($v = c z$) for all velocities.  The
atomic hydrogen (H{\sc i}) column densities are extremely high in the bridge
region, reaching about $3 \times 10^{21}$ cm$^{-2}$.  Two H{\sc i} peaks,
H{\sc i}$_{\rm E}$ and H{\sc i}$_{\rm W}$,
are also present in the outer parts of the common H{\sc i} envelope
and are probably tidal features.  The western H{\sc i} peak coincides
with a blue object.  The eastern H{\sc i} peak has no obvious optical
counterpart but is close to the optical tidal tail.

\citet{Condon02} suggests
that UGC 813 and UGC 816 are undergoing their first collision (some 40
to 50 Myrs ago) and are now separating at about 400 -- 500 km s$^{-1}$.
Contrary to the UGC 12914/5 system, the UGC 813/6 disks are rotating in
the same direction.  CGCG 551-011 is also a spiral galaxy but has been
stripped of most of its H{\sc i} and has a strong compact radio continuum
source.  Without numerical simulations, we cannot determine whether
the tidal features could have been generated by a collision between UGC~816
and CGCG~551-011 some 200 Myr ago, stripping CGCG~551-011 of most of its
H{\sc i} and feeding the central source.

After describing the observations of the UGC~813/6 system, we present
the CO spectra and compare these with the H{\sc i} data from \citet{Condon02}.
Using the information available for both the UGC 12914/5 and UGC 813/6
pairs, we then show that molecular clouds can collide and drive large
amounts of dense gas out of the spiral disks, despite earlier assumptions
to the contrary.  The differences with respect to previous work are explained
and a general scenario proposed for how the gas, particularly the
dense gas, gets into the bridge region for this type of collision.
Taking H$_0 = 70$ km s$^{-1}$ Mpc$^{-1}$, we assume a distance to the
UGC 813/6 group of 75 Mpc.

\section {Observations}

The UGC~813/6 bridge, galaxy centers, and the outlying H{\sc i} maxima
were observed in the CO(1$\rightarrow$0) and CO(2$\rightarrow$1) to
search for molecular gas.
The observations were carried out with the 30 meter millimeter-wave
telescope on Pico Veleta (Spain) run by the Institut de RadioAstronomie
Millim\'etrique (IRAM) in August 2003.  The CO(1$\rightarrow$0) and
CO(2$\rightarrow$1) transitions at 115 and 230 GHz respectively were
observed simultaneously and in both polarizations.  A bandwidth of
over 1300 km s$^{-1}$ was available in both transitions using the two
$512 \times 1$ MHz filterbanks at 3mm and the two $256 \times 4$ MHz
filterbanks at 1mm.  The band was centered around $cz = 5200$ km s$^{-1}$,
where $z$ is the redshift, corresponding to redshifted frequencies of
113.3059 and 226.6074 GHz.

\begin{figure*}[t]
\begin{center}
\includegraphics[angle=270,width=18cm]{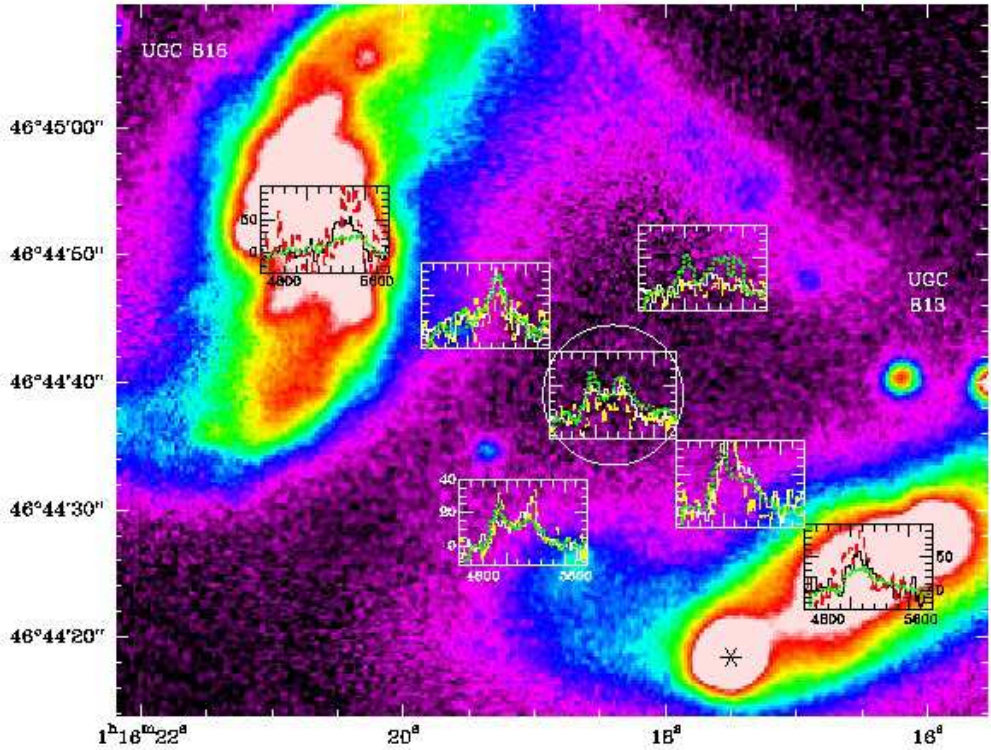}
\caption{CO(1$\rightarrow$0), CO(2$\rightarrow$1), and H{\sc i} spectra
at the positions observed in the UGC 813/6 galaxies and bridge, using
respectively black solid (white solid in the bridge), red dashed
(yellow dashed in the bridge),
and green dotted lines.  The circle indicates the size of the CO(2--1) beam.
UGC 813 is the edge-on spiral to the lower right and UGC 816 is to the left,
less inclined and at a higher velocity.  The asterisk marks a bright
foreground star.  For the CO
spectra, the intensity scale is main beam antenna temperature in mK
(-30 to 100 mK for the galaxy centers and -12 to 40 mK for the bridge
positions).  The H{\sc i} spectra are plotted such that equal
CO(1$\rightarrow$0) and H{\sc i} intensity correspond to equal H-atom
column densities in H$_2$ and in H{\sc i}.  It is thus apparent that
in the bridge, the H{\sc i} and H$_2$ column densities are similar,
assuming the $\ratio$ factor is correct.  The underlying image is an
R band image taken with the Instituto de Astrof\'{\i}sica de Andaluc\'{\i}a
(IAA) 1.5--m telescope on Pico Veleta.  }
\end{center}
\end{figure*}

System temperatures were typically 200 -- 300 K at 3mm and twice as high
for the CO(2--1) transition (T$_A^*$ scale).  The forward (main beam)
efficiencies at Pico Veleta are currently estimated at 0.95 (0.74) at
115 GHz and 0.91 (0.54) at 230 GHz.  At the redshifted frequencies
the half-power beamwidths are 22$''$ and 11$''$.
All observations were done in wobbler-switching mode, usually with a
throw of 100$''$ but sometimes more or less depending on the position
observed, in order to be sure not to have emission in the reference beam.
Pointing was checked on the nearby quasar 0133+476 every
60 -- 80 minutes with pointing errors usually less than 3$''$.

Data reduction was straightforward.  After eliminating the few obviously
bad spectra or bad channels, the spectra for each position and each
transition were
summed.  Only zero-order baselines ({\it i.e.} continuum levels) were
subtracted to obtain the final spectra.  Where no CO line was obvious,
the line windows were based on the H{\sc i} spectra.

\section {Results}

The centers of both galaxies and 4 of the 5 bridge regions were clearly
detected in both transitions.  In Fig. 1 we show the CO(1--0) and CO(2--1)
spectra for the galaxy centers and bridge along with the H{\sc i} spectra
at those positions at 16$''$ resolution, intermediate between the
CO(1--0) and CO(2--1) angular resolutions.  The northernmost bridge
position shows positive flux at the appropriate velocities.  No line is
obvious in CO(2--1) although it covers an optically brighter region
(a tidal arm) than
the bridge center (Fig. 2).  All of the pointing centers are shown in
Fig. 2.  In Fig. 3 we show the CO(1--0), CO(2--1), and H{\sc i} spectra
of the eastern and western H{\sc i} peaks.  No CO emission was detected
despite the strong H{\sc i}.

Comparison of the CO(1--0) observations at 22$''$ (this paper) and
55$''$ \citep{Zhu99}
suggests that in UGC 813 the molecular gas is more centrally
concentrated than in UGC~816 because in UGC 813 the line is much
brighter when observed at high resolution.

\begin{figure*}[t]
\begin{center}
\includegraphics[angle=270,width=18cm]{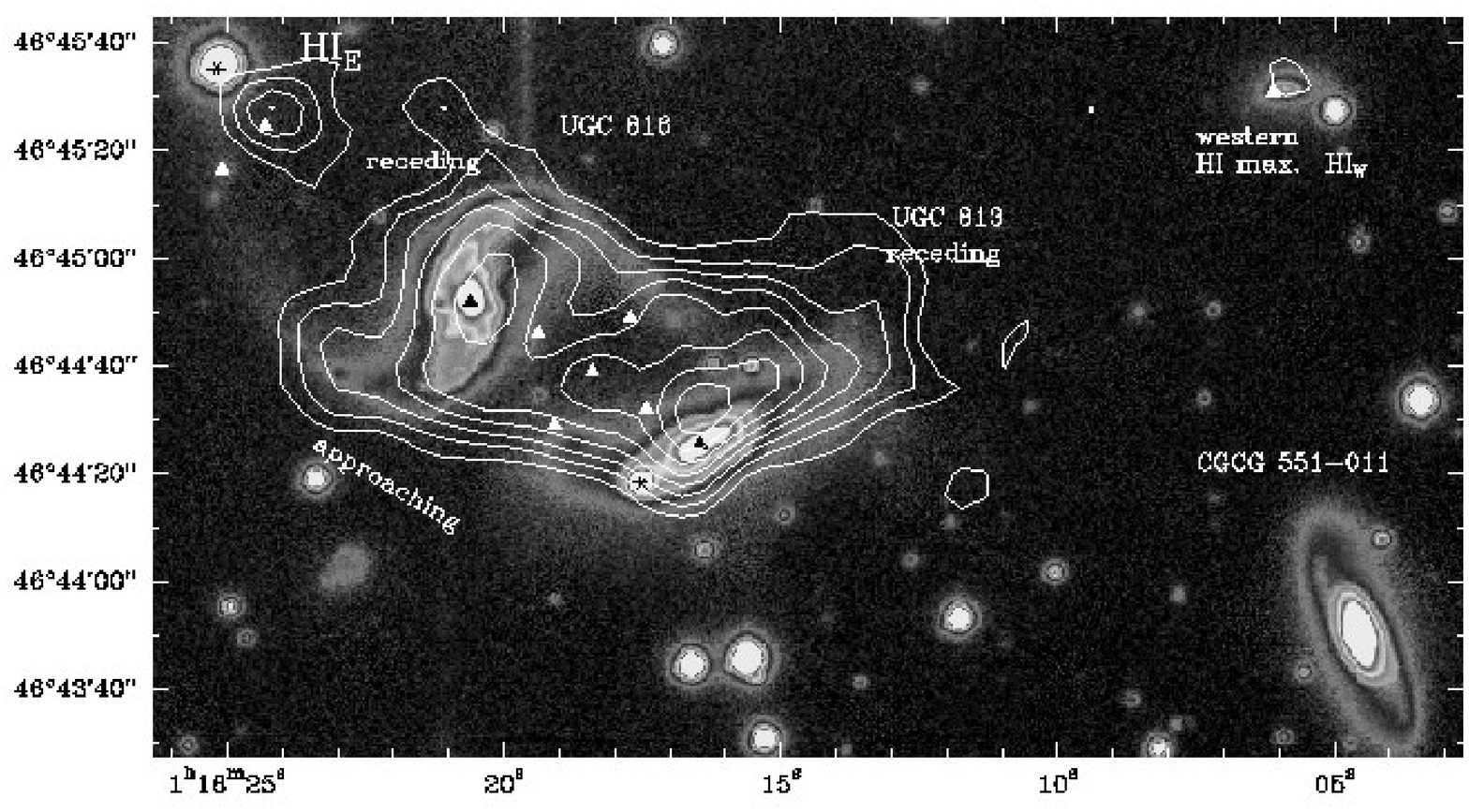}
\caption{R band image taken with the IAA 1.5meter telescope of the
entire UGC 813/6 system with
H{\sc i} column density contours from \citet{Condon02}.
The centers of UGC 813 and 816 are marked by black triangles.
The bridge positions observed are the five white triangles between the
galaxies.  These are the positions whose spectra are shown in Fig. 1.
The triangles to the extreme West and East, slightly north of UGC 816,
mark the positions of the H{\sc i} peaks whose spectra are shown in Fig. 3.
Asterisks mark the positions of some galactic stars which could be confused with
other features.  The spiral galaxy seen nearly edge-on in the lower
right is CGCG~551-011.}
\end{center}
\end{figure*}

\begin{table*}
\begin{center}
\begin{tabular}{llllllll}
Source & RA & Dec & offset & I$_{\rm CO(1-0)}$
 & I$_{\rm CO(2-1)}$ & N(H$_2$) & N(H{\sc i}) \\
& (J2000) & (J2000) & arcsec & K km s$^{-1}$ & K km s$^{-1}$ & 10$^{20}$ cm$^{-2}$& 10$^{20}$ cm$^{-2}$\\
\hline
UGC 813 & 01:16:16.45 & 46:44:25 & (0,0) & 12.2$\pm 0.9$ & 10.3$\pm 1.3$ & 24 & 38 \\
UGC 816 & 01:16:20.6 & 46:44:52 & (0,0) & 11.1$\pm 0.6$ & 13.9 $\pm 2.1$ & 22 & 37 \\
Bridge  & 01:16:18.4 & 46:44:39 & (0,0) & 7.0$\pm 0.2$ & 3.7$\pm 0.4$ & 14 & 35 \\
Bridge  &&& (7,-10) & 6.6$\pm 0.4$ & 7.4$\pm 0.5$ & 13 & 27 \\
Bridge  &&& (-7,10) & 2.5$\pm 0.3$ & 0.7$\pm 0.4$ & 5 & 32\\
Bridge  &&& (10,7) & 6.7$\pm 0.4$ & 4.5$\pm 0.7$ & 13 & 30\\
Bridge  &&& (-10,-7) & 10.5$\pm 0.5$ & 8.3$\pm 0.6$ & 21 & 35 \\
H{\sc i}$_{\rm E}$ & 01:16:24.31 & 46:45:24 & (0,0) & 0$\pm 0.14$ & 0$\pm 0.4$ & $\la 1$ & 22 \\
H{\sc i}$_{\rm E}$&&& (8,-8) & 0$\pm 0.10$ & 0$\pm 0.2$ & $\la 0.6$ & 8\\
H{\sc i}$_{\rm W}$ & 01:16:06.1 & 46:45:31 & (0,0) & 0$\pm 0.08$ & 0$\pm 0.2$ & $\la 0.5$ & 10 \\
\end{tabular}
\caption[]{Complete list of positions observed in CO, corresponding to the
triangles in Figure 2.  The offsets are in arseconds with respect to
the (0,0) position of the source.
Columns 3 and 4 give the CO(1--0) and CO(2--1)
integrated intensities and cols. 5 and 6 the H$_2$ and H{\sc i}
column densities.
The CO line window for the "H{\sc i}$_{\rm E}$" positions is 5370 to
5470 km s$^{-1}$ and 5170 to 5270 km s$^{-1}$ for the "H{\sc i}$_{\rm W}$"
position, both based on the H{\sc i}
spectra shown in Fig. 3.  Uncertainties are 1$\sigma$ based on the rms noise
and the line width.  Upper limits to N(H$_2$) are $3\sigma$.}
\end{center}
\end{table*}

In Table 1 we give the CO intensities for each transition and the
corresponding H$_2$ and H{\sc i} column densities for each of the positions
observed.  While the CO emission from the galaxies is stronger than in the
bridge, the CO emission from the bridge is at least that of the Milky Way.
For a CO -- H$_2$ conversion factor of $\ratio = 2 \times 10^{20}$ molecules
cm$^{-2}$ (K km s$^{-1}$)$^{-1}$ \citep[e.g.][]{Dickman86}, the bridge is
about half molecular and half atomic gas -- some 2 $\times 10^9$M$_\odot$
of each.  This is based on the central bridge position alone.  For the other
positions, it is difficult to separate the emission from the disks from
that of the bridge, such that the above represents a lower limit to
the true bridge CO emission.

\begin{figure}[t]
\begin{center}
\includegraphics[angle=270,width=8.8cm]{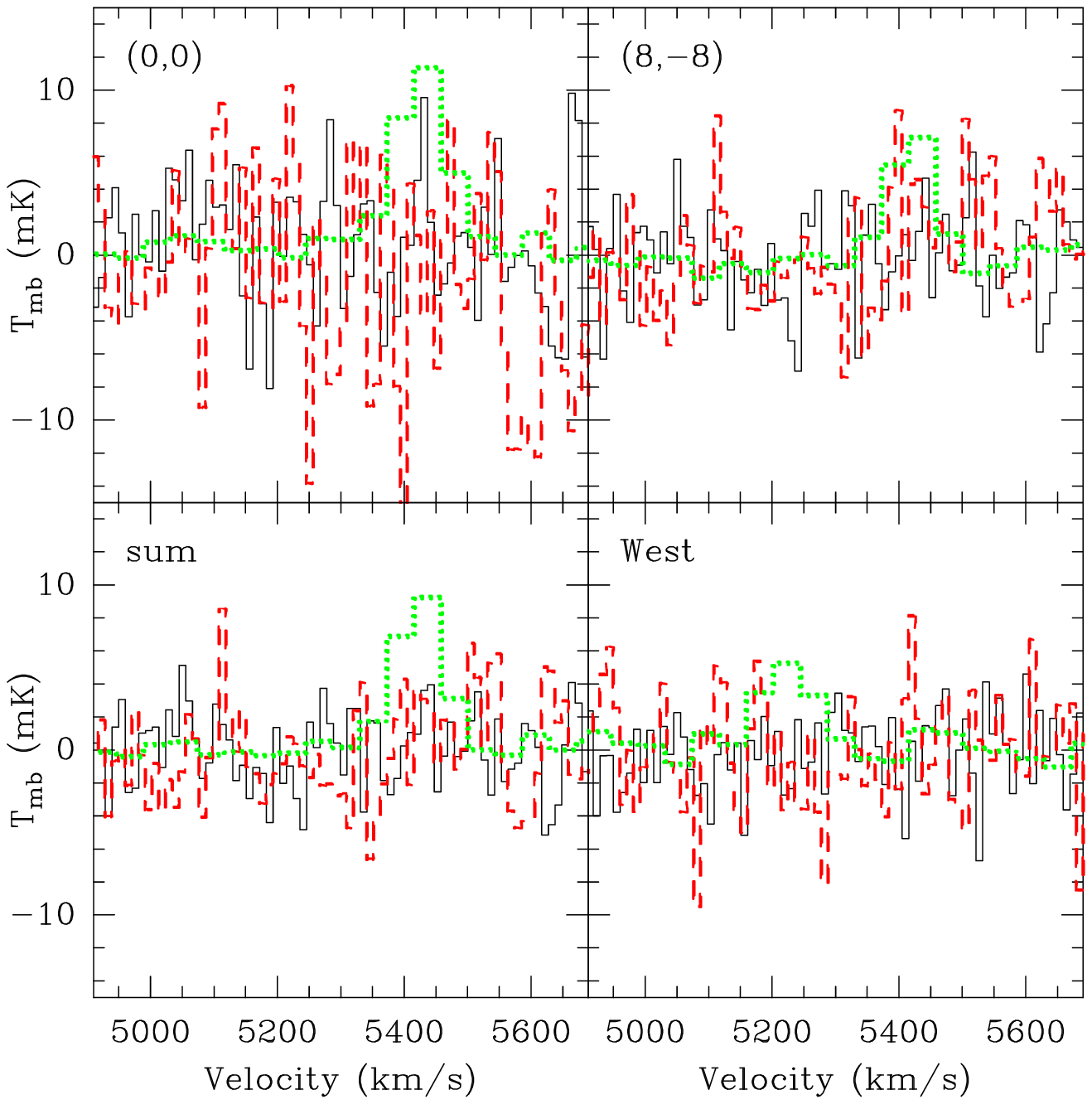}
\caption{CO(1--0), CO(2--1) and H{\sc i} spectra of the "TDG candidate"
positions.  The brightness scale is for the CO(1--0) line and should be doubled
for the CO(2--1) line ({\it i.e.} the  CO(2--1) spectra have been divided by two).
The H{\sc i} scale has been divided by 4 such that if the column density
of molecular gas were 25\% of the atomic gas column density, then the spectra
would be of the same height.  It is immediately obvious that the CO intensities
indicate low H$_2$ column densities compared to H{\sc i}.
The lower left panel shows the sum of the two eastern positions
shown in the top two panels.  The positions are shown in Fig. 2 where
they are indicated by triangles. }
\end{center}
\end{figure}

As can be readily seen from Fig. 1, the molecular and atomic gas line
profiles are very similar.  Furthermore, assuming the $\ratio$ factor used
is appropriate, the hydrogen column densities are also similar in
the bridge region (the H{\sc i} spectra in Fig. 1 are scaled so that
the same line area as in CO represents the same H-atom column density).
The molecular gas dominates in the galactic centers and in the densest
parts of the bridge.  This is obviously not true of tidal tails, which
preferentially bring material out of the H{\sc i} - dominated outer regions
of spirals.  The CO line intensities given in Table 1 for the bridge positions
are unheard of for tidal material.  The two H{\sc i} peaks, at about 1-2 arcmin
from the bridge and marked as H{\sc i}$_{\rm E}$ and H{\sc i}$_{\rm W}$,
appear to be at the ends of tidal arms or tails and indeed the H{\sc i}
emission is strong but no CO has been detected there so far (Fig. 3).
The H$_2$/H{\sc i} mass ratio in the tidal material is less than 10\%,
compatible with the less evolved Tidal Dwarf Galaxies (TDG) for which CO
detections have been obtained \citep{Braine_tdg2}.  Figure 4 presents
an H$\alpha$ image of the UGC~813/6 system with H{\sc i} contours superposed.
No star formation is observed near the northeastern H{\sc i} peak
(H{\sc i}$_{\rm E}$), which is clearly not, or at least not yet, a TDG.

To further distinguish between bridge and tidal features, we plot the
synchrotron emission \citep{Condon02} over the R band image in Fig. 5.
Just as in the Taffy galaxies, the CO emission follows the synchrotron
more than any tidal features.  Strong extra-disk synchrotron emission
is one of the characterizing features of ISM-ISM collisions and not of
tidal arms or tails.

In the Taffy galaxies (UGC 12914/5), the CO/H{\sc i} ratio in the bridge is
higher.  The CO and H{\sc i} line profiles in the UGC 12914/5 bridge are
quite different; this is, however, probably a projection effect because we
see the H{\sc i} from UGC 12914 as if it were in the bridge while it is
actually far behind \citep[see ][]{Braine03}.
UGC 12914 and 12915 are generally bigger and more gas-rich
than the UGC 813/6 system.  As we will see, this is important for the amount
of molecular gas that can be stripped through cloud -- cloud collisions.

\section {Why is H$_2$ found in the bridges?}

In the following we will address the question of the origin of the H$_2$
found in the bridge between UGC~813 and UGC~816.  As this galaxy pair is
very similar to the Taffy galaxies UGC~12914/5 and we also found abundant
molecular gas there, we will attempt to provide a more general explanation
which applies to any Taffy--type interaction.  We will assume
that the $\ratio$ factor adopted here is valid for the bridge molecular gas.
\citet{Gao03} adopt a higher $\ratio$ factor in their work on UGC 12914/5;
\citet{Braine03} discuss why the factor they adopt is probably too high.
The first question is perhaps
whether the molecular gas could form from the atomic gas in the bridge.

\subsection{Could the H$_2$ form out of the H{\sc i} in the bridge?}

The H{\sc i} column density is typically $2 - 4 \times 10^{21}$ cm$^{-2}$
in the bridges.  If the depth of the bridge is similar to its extent
perpendicular to
the collision direction, then the neutral gas is spread over some 10 -- 20 kpc.
The average volume density is then $\la 0.1$.  \citet{Braine_tdg2} estimate
that $10^7/n$ years ({\it i.e.} $\ga 10^8$ yrs in the bridges) is necessary
to convert 20\% of the H{\sc i} into H$_2$.  Once some of the H{\sc i} has
become H$_2$ the process becomes less efficient.  While local density
enhancements could allow some H$_2$ formation from the atomic gas, it is
unlikely that the large quantities of H$_2$ observed could have
formed from H{\sc i}.  H$_2$ is in fact the {\it dominant} phase in the
Taffy bridge and is roughly half of the UGC~813/6 bridge ISM.
If the grains are warm then the
ability of H-atoms to form H$_2$ on the grain surfaces is greatly reduced
\citep{Hollenbach71a}.  Furthermore, the efficiency of
H{\sc i} to H$_2$ conversion is obviously reduced if grains are destroyed
in the strong shocks generated in cloud collisions (whether in atomic or
molecular clouds).

There is a further reason for why clouds which are not initially dense
will not be able to contract.  In both systems, and particularly the Taffy
galaxies, the spirals are massive (several $10^{11}$ M$_\odot$ each from
C93 and \citet{Condon02}).  As a result, the tidal forces they exert
on clouds are stronger than in the Milky Way.  Following \citet{Bania86},
for clouds to resist against tidal shear, their density must be
$$ n \ga 200 \, ({{3 {\rm \, kpc}} \over{R}})^2 {\rm nucleons \, \, cm}^{-3}$$
where R is the distance from the cloud to the galactic centers.
Being able to resist tidal forces is not sufficient for cloud survival
but is certainly required.  Thus, clouds which are not sufficiently
dense after collision to form H$_2$ rapidly may not be dense enough to
survive to form H$_2$ subsequently in the bridge.

\subsection{GMC collisions as part of galaxy collisions}

The observations show that when spiral galaxies physically collide, with
the inner parts of their disks passing through each other,
more than $10^9$ M$_\odot$ of molecular gas is found in the resulting
bridge region between the two galaxies.  While tidal forces can drive
gas out of spiral galaxies, they act on stars in the same way, such
that it is difficult to imagine how a bridge with so much gas but no stars
could have been formed by tidal forces alone.
It has commonly been accepted \citep[e.g.][]{Jog92} that Giant Molecular
Clouds (GMCs) do not collide whereas the more diffuse clouds of atomic gas do.
The goal of this section is to demonstrate that in gas-rich but not
exceptional spiral galaxies GMC-GMC collisions occur and are likely
the source of the large quantities (1 -- 10 $\times 10^9$M$_\odot$)
of molecular gas present in the bridges.

\begin{figure*}[t]
\begin{center}
\includegraphics[angle=270,width=18cm]{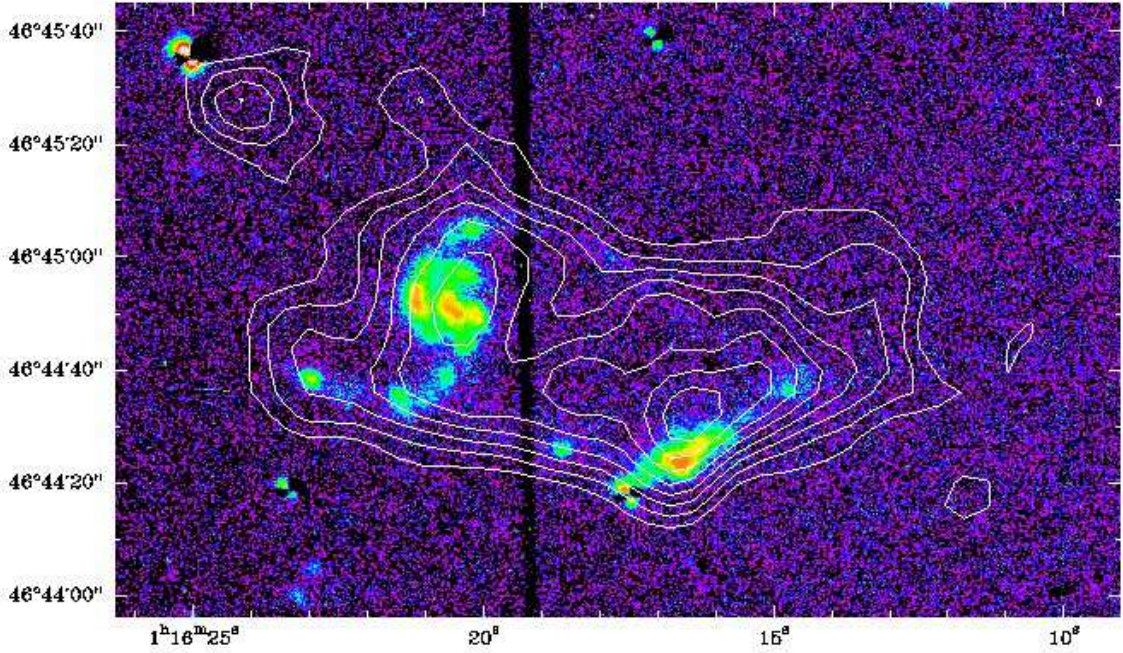}
\caption{H$\alpha$ image after broadband continuum subtraction taken
with the San Pedro M\'artir, Baja California (Mexico), 2.12--m telescope
with the same H{\sc i} column density contours as in Fig. 2.
The dark vertical
stripe through the bridge is due to the bright star above UGC~816.
A fairly bright H{\sc ii} region is seen in the southern part of
the bridge.  No H$\alpha$ emission is observed near the NE H{\sc i} peak.}
\end{center}
\end{figure*}

\begin{figure*}[t]
\begin{center}
\includegraphics[angle=270,width=18cm]{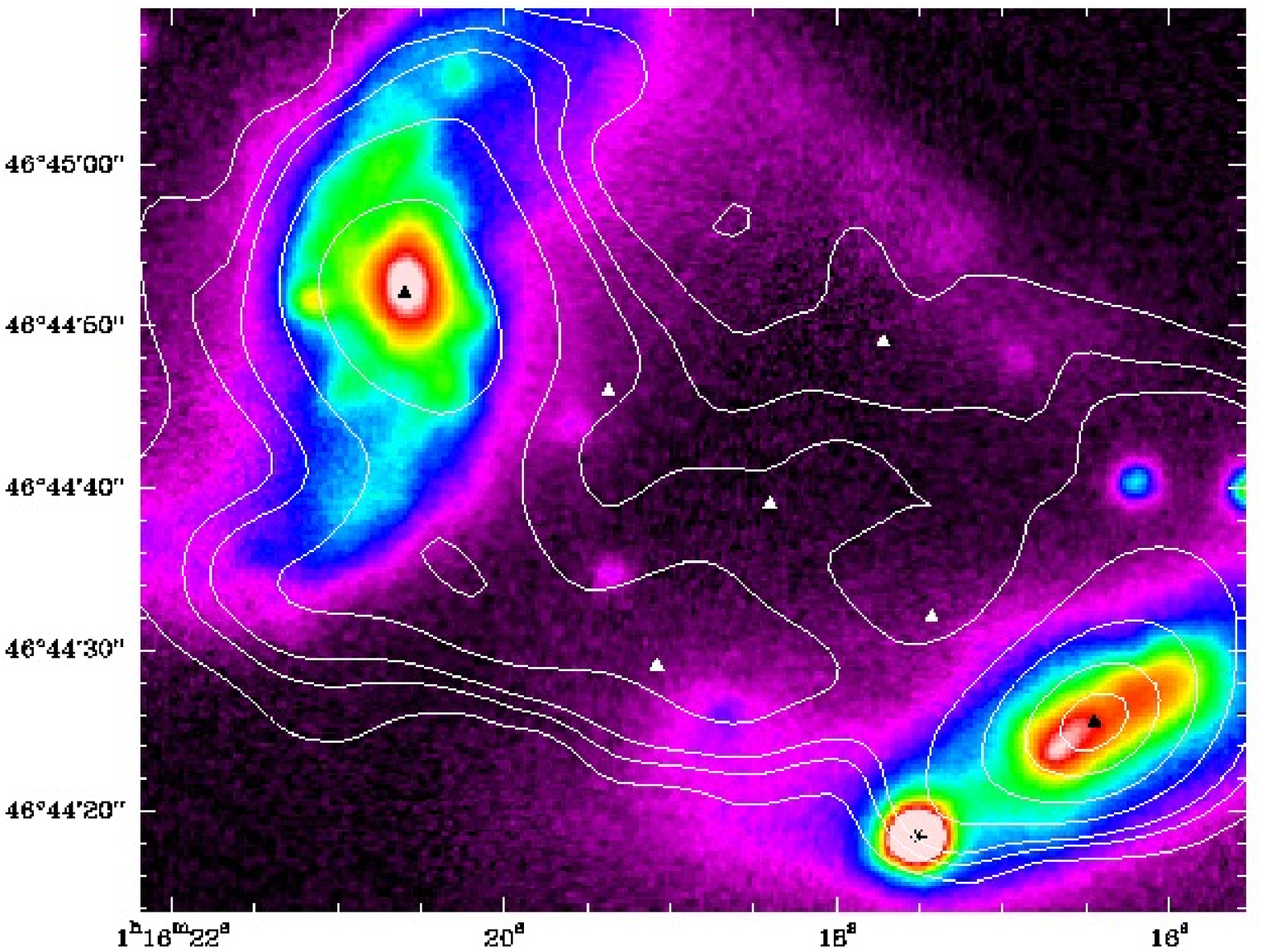}
\caption{R band image with 1.4 GHz synchotron emission contours
superposed.  Triangles indicate the positions observed in CO and the
asterisk a star.  While the optical emission in the NW bridge position
(near the northern tidal arm) is stronger than in the others, the CO
emission is much weaker there. }
\end{center}
\end{figure*}

In the calculations we will suppose that molecular clouds are spherical
and obey the so-called \citet{Larson81} relation with an average column
density of N(H$_2$) $= 10^{22}$ cm$^{-2}$.  This is typical and corresponds
exactly to what \citet{Jog92} assume.
We define the following quantities: \\
$A_{cl}$, the projected cloud area, equal to $\pi R_{cl}^2$ for
identical spherical clouds where $R_{cl}$ is the cloud radius; \\
$n_{cl}$, the number density of such clouds; \\
$M_{cl}$, the mass of a cloud of radius $R_{cl}$; \\
$z_{\rm H_2}$, the half thickness of the layer of molecular clouds;\\
$\Sigma_{\rm H_2}$, the face-on H$_2$ surface density of the spiral galaxies,
expressed in M$_\odot$ pc$^{-2}$, typically averaged over a telescope beam;\\
$\Sigma_{cl}$, the H$_2$ surface density of a cloud which is simply
160 M$_\odot$ pc$^{-2}$ for N(H$_2$)  $= 10^{22}$ cm$^{-2}$; \\
and the velocities $V_{\perp}$, taken to be the encounter speed perpendicular
to the galactic planes and $V_{\parallel}$, the relative cloud velocities
in the planes (which are assumed parallel as in the Taffy galaxies).
The link to extragalactic observables is provided by the integrated
CO(1$\rightarrow$0) line intensity I$_{\rm CO}$, expressed in K km s$^{-1}$,
via the conversion factor taken to be $\ratio = 2 \times 10^{20}$ H$_2$
cm$^{-2}$ per K km s$^{-1}$.

The mean free path $\lambda_{\rm mfp}$ of a spherical particle is
$\lambda_{\rm mfp} = 1/(4 A_{cl} n_{cl})$.
For a gas of particles in random motion, the mean free path is actually
$\sqrt{2}$ shorter.  The factor 4 comes from the fact that clouds collide
if they pass within 2$R_{cl}$ of each other.  The number density of clouds is
$$n_{cl} = {{\Sigma_{\rm H_2}} \over {2 z_{\rm H_2}}} \times {{1}
\over {M_{cl}}},
\, \, \, {\rm where} \, \, \, M_{cl} = \Sigma_{cl} \times A_{cl}$$
such that the mean free path is simply
$$\lambda_{\rm mfp} = {{2 z_{\rm H_2} \Sigma_{cl}} \over {4\Sigma_{\rm H_2}}} $$
such that when $\Sigma_{\rm H_2} = \Sigma_{cl}$, the mean free path
of a cloud is 1/4 of the thickness of the molecular cloud layer
$2 z_{\rm H_2}$.  This is in some ways a trivial result because
$\Sigma_{\rm H_2} = \Sigma_{cl}$ means that the galactic planes are
completely covered by a layer of molecular clouds so it seems obvious
that they should collide when the disks pass through each other.
CO observations show that many spirals have disk CO intensities of order
I$_{\rm CO} \ga 20$ K km s$^{-1}$, corresponding to $\Sigma_{\rm H_2} \ga
65$ M$_\odot$ pc$^{-2}$, such that the mean free path is less than the
thickness of the molecular cloud layer.

The mean free path formula above gives collisions even when only parts
of the spherical clouds overlap.
However, when there is a transverse velocity, such as in the
Taffy galaxies where the counter-rotation yields high encounter speeds
parallel to the galactic planes, then the path of a cloud in the molecular
layer of the other system is increased from 2 $z_{\rm H_2}$ to 2
$z_{\rm H_2} \sqrt{V_{\parallel}^2 + V_{\perp}^2} / V_{\perp}$.
When the parallel and perpendicular velocities are roughly equal, this
lengthens the path and increases the probability of cloud collisions by
$\sqrt{2}$.  While the magnetic field energy in a cloud is greatly less
than the kinetic energy of the collision, magnetic fields are nonetheless
quite strong in dense gas, with the field strength increasing as
$\sqrt{\rm density}$, reaching about 100 $\mu$ G \citep{Crutcher99}.
As such, they could be expected to increase the effective cloud cross-section
somewhat.  If the galactic planes are not parallel and/or if the collision
is not face-on then the cloud paths within the galactic planes are
increased.

The above discussion only considers collisions between molecular
clouds, showing that rather than being rare, GMC - GMC collisions on
large scales are inevitable in collisions of gas-rich spirals.
After (inelastic) collision, the resulting cloud and possibly cloud fragments
will have a velocity intermediate between the velocities of the galactic
disks.  They are thus left behind in the bridge region between the separating
galaxies.  Whether or not the clouds fall back onto one of
the galaxies depends on their velocity with respect to the galaxies and
thus the collision velocity.

\subsection{Why previous results were different}

We choose to compare here with the work of \citet{Jog92} who explored this
question and have been cited for why GMCs should {\it not} collide.
Furthermore, the cloud properties they assume are in accord with our
hypotheses -- the only significant number, the average H$_2$ column density
(or projected H$_2$ surface mass density) is the same as our adopted value.
\citet{Jog92} define the mean free path by introducing a volume filling factor
for molecular clouds, which is the cloud volume multiplied by the number of
clouds per unit volume in the molecular layer of a spiral galaxy.  Hence
\begin{eqnarray}
\nonumber
f & = & 4 \pi \, R_{cl}^3 \, n_{cl} /3 \\
\nonumber
 & = & (4 \pi R_{cl}^3 /3) \times \Sigma_{\rm H_2}/(2 z_{\rm H_2} \Sigma_{cl} A_{cl}) \\
\nonumber
 & = & (4 R_{cl} / 3) \times \Sigma_{\rm H_2} / (2 z_{\rm H_2} \Sigma_{cl})
\end{eqnarray}
after making the substitutions using the formulae above for identical spherical
clouds (the hypothesis adopted by \citet{Jog92}).  Their mean free path is then
$$ \lambda_{\rm mfp} = 2 R_{cl} / 3 f
 = 2 R_{cl} / 3 \times 1 / (4 \pi / 3) R_{cl}^3 \, n_{cl}
 = 1/2 A_{cl} n_{cl} $$
which is twice the value we use.  This may be to ensure that only collisions
with comparable molecular column densities are included, such that grazing
encounters are entirely excluded.  This is not the origin of the discrepancy.

\citet{Jog92} specify the volume filling factor $f$ to be
$f_{\rm GMC} = 0.01$ which we now express in terms closer to observables.
Using the above definitions,
\begin{eqnarray}
\nonumber
\Sigma_{\rm H_2} & = & f \times 2 z_{\rm H_2} \Sigma_{cl} A_{cl} \times
3 / (4 \pi R_{cl}^3) \\
\nonumber
 & \approx & {{f} \over {0.01}} \times {{2 z_{\rm H_2}} \over {130{\rm pc}}}
 \times {{\Sigma_{cl}} \over {R_{cl}}} {\rm M}_\odot {\rm pc}^{-2} \\
\nonumber
 & \approx & 6 {\rm M}_\odot {\rm pc}^{-2}
\end{eqnarray}
for  $2 z_{\rm H_2} = 130$pc, $\Sigma_{cl} = 160 {\rm M}_\odot {\rm pc}^{-2}$,
and $R_{cl} = 25$pc, as in \citet{Jog92}.  $6 {\rm M}_\odot {\rm pc}^{-2}$
corresponds to I$_{\rm CO} = 2$ K km s$^{-1}$, which is weak for the
inner parts of a spiral
galaxy.  {\it This is the origin of the discrepancy: the \citet{Jog92}
calculations assume a very low molecular gas surface density.}
The value they assume is roughly the molecular gas surface density of
the Milky Way averaged over the entire optical disk.  In the molecular
ring of our Galaxy the H$_2$ surface density is much higher.  This is why
grazing encounters should not be efficient at generating GMC-GMC collisions
but collisions of the inner parts of spirals will be.

\section{ The collision scenario}

Unfortunately, no simulations of colliding neutral clouds, especially with
magnetic fields, are available.  \citet{Harwit87} explore analytically
what happens when two (big) disks of dense gas collide -- we largely follow
their scenario.  The crossing or collision time of two clouds is
$$t = D_{eff}/V  = 10^5 \, {{\sqrt{A_{cl}}} \over {50{\rm pc}}}
\times {{500 \kms} \over {V}} \, \, \, \, \, \, {\rm years.}$$
As two clouds collide, whether originally atomic or molecular, the colliding
parts are instantly ionized and reach temperatures of $T \ga 10^6$ K
(depending on original collision velocity).  As the two clouds continue
to move into each other, the central region will be compressed and will cool
very rapidly.  A similar scenario is proposed in the case of interstellar
shocks in Fig. 1 of \citet{Hollenbach89}.
The pressure in the central region is very high so the gas
will spread out perpendicular to the collision direction, as in the
\citet{Mair88} simulations.  How much gas flows to the side is highly
dependent on the geometry of the clouds and of the collision.  With the
exception of gas flow towards the exterior, the collisions are concisely
described by Figure 4 and Sect. 8 of \citet{Harwit87}.  The \citet{Struck97}
simulations also suggest that ISM-ISM collisions can bring large amounts
of gas into the bridge and that the gas is dense enough to cool, although
he does not go into the details of cooling or the cloud collisions themselves.

The hot gas ($10^6 \la T \la 10^7$ K) cools very rapidly
\citep{Gaetz83,Harwit87} down to below $10^4$ K.  The cooling
timescale, only a few years for molecular cloud collisions, is
much shorter than the collision timescale.  Thus, the material at
the barycenter of the collision will very quickly become cool and
neutral, provided it is sufficiently dense.
\citet{Hollenbach89} estimate the column density necessary to form
H$_2$ to be $N_{1/2} \approx 4 \times 10^{21}$ cm$^{-2}$ where the
subscript 1/2 indicates changing half of the H{\sc i} into H$_2$.
The longest time scale is also the H$_2$ formation time scale.
All of the time scales vary as the inverse of the density so
collisions of dense material have the best chances of producing
predominantly cool molecular gas at the end of the collision. This
means that the molecular gas we observe in the bridge region left
the galactic disks in molecular form.

Most collisions will obviously not be head-on collisions between
identical clouds.  Without detailed simulations of neutral cloud
collisions, however, it is difficult to estimate, for each
geometry and possible cloud type (atomic, molecular), the fraction
of the gas post-collision which will be in atomic, molecular, or
ionized form.  It is likely that head-on collisions of large
atomic clouds will also produce (after ionization and cooling) a
high fraction of H$_2$.  Given that ($i$) collisions between
molecular clouds are not only possible but inevitable on large
scales once the {\it surface} filling factor ($\sim \Sigma_{\rm
H_2}/\Sigma_{cl}$) of the colliding spirals is greater than about
0.2 ($\lambda_{\rm mfp} \sim$ thickness of molecular gas layer)
and ($ii$) that collisions between dense clouds are most likely to
generate cool dense (molecular) gas post-collision, it is
reasonable to expect much of the H$_2$ in the bridge regions to be
due to collisions of molecular clouds.

 Observationally, if H{\sc i} -- H{\sc i} cloud
collisions were efficient in creating H$_2$, then the H{\sc i}/H$_2$ ratio in
the Taffy bridge would be relatively constant.  However, the H$_2$ in the
bridge decreases rather sharply from the center outward, just like the
centrally concentrated CO emission in spirals, and the H{\sc i} is extended
both in the bridge and in the spirals.  The most likely explanation is that
molecular cloud collisions generate much of the observed molecular gas
in the bridges.

\section{Star Formation in the bridges?}

With such large quantities of molecular gas present, star formation
rates of 1 -- 10 $M_\odot$ per year are expected in the bridges, if
the gas is in a ``normal'' state, capable of forming stars as in spirals.
From the H$\alpha$ and CO emission of spirals, a star formation rate (SFR)
of about 1$M_\odot$ per year is estimated per $10^9 M_\odot$ of molecular gas
\citep[][ Sect. 3]{Kennicutt98b,Braine_tdg2}.  The bridges are estimated
to be about 20 -- 30 Myr and 40 -- 50 Myr old for the UGC 12914/5 (Taffy)
and UGC 813/6 systems \citep{Condon93,Condon02}.  Such times are well
above the collapse times for cloud cores.

From the CO emission from the bridges, the SFRs expected in
the UGC 12914/5 and UGC 813/6 bridges are about 9 and 2 $M_\odot$yr$^{-1}$.
\citet{Jarrett99} find that with the exception of the single giant
H{\sc ii} region near UGC 12915, the UGC 12914/5 bridge is devoid of
stars and star formation.  From the slit spectrum of the UGC 12914/5
bridge presented by \citet{Braine03}, there is no
sign of star formation (except in the H{\sc ii} region) although
the molecular gas surface density is nearly constant \citep{Braine03,Gao03}
throughout the bridge.  A single discrete H{\sc ii} region is observed
in the UGC 813/6 bridge (Fig. 4).
The luminosity expected from bridges with several $M_\odot$yr$^{-1}$ of
star formation is not observed.  Because any stars formed in the bridge
are necessarily quite young, could they still be dust-enshrouded,
such that we would not see their light?

From the H$\alpha$ observations \citep{Bushouse87,Bushouse90} of the
Taffy pair, we estimate that the SFR is about 0.1 -- 0.3 $M_\odot$yr$^{-1}$.
From the Far-IR (IRAS) luminosity of the pair, about $4 \times 10^{10}$L$_\odot$
between 42 and 122$\mu$m, the SFR is between 5 and 10 $M_\odot$yr$^{-1}$,
more than half of which is in UGC 12915. In the Taffy galaxies there is
a lack of FIR emission in the bridge \citep{Zink00} but dust, perhaps
only small grains, is detected via its Mid-IR emission.
The flux in the bridge at 15$\mu$m measured by \citet{Jarrett99} with
ISOCAM is about 5\% of that measured in each of the galaxies themselves.
 This is in rough agreement with
the H$\alpha$ based SFR so we estimate that the bridge SFR is between
0.1 and 0.4 $M_\odot$yr$^{-1}$.  For the Taffy system, this is less
than 10\% of what is expected based on the CO luminosity.

The cloud collisions and associated shocks are expected to destroy dust
grains to a large extent.  No observations relevant to this are available
for the UGC 813/6 system.  The small amount of dust in the
Taffy bridge \citep{Zink00} means that it is unlikely
that star formation is hidden through extinction by dust.  The dip in the FIR
emission in the bridge and the low SFR calculated from the MIR emission
indicates that less star formation is occurring in the Taffy bridge than
would be expected
given the strong CO emission.  As discussed in \citet{Braine03}, the
$\ratio$ ratio could be lower by a factor of a few.  This would reduce
the amount of ``missing'' star formation but even allowing for the
uncertainty in the $\ratio$ conversion, the SFR still appears low.

In the UGC~813/6 system, an H{\sc ii} region is present in the bridge but
contributes less than 2\% of the H$\alpha$ luminosity of the system.
The H$\alpha$ emission surely suffers from extinction in the galaxies,
particularly in UGC~813 which is seen edge-on.
Although the pair has not been completely mapped in CO, we estimate that
the bridge contributes between 12 and 30\% of the global CO luminosity.
Thus, again, the SFR is only some 10\% of what would be expected in
``normal'' molecular gas.  {\it Our tentative
conclusion is that star formation in the bridge material has been retarded
or even suppressed by the recent collision.}

In galaxies, stars form through the collapse of dense ``pre-stellar'' cores
whose densities are above $10^5$ cm$^{-3}$ and column densities well above
$10^{22}$ cm$^{-2}$.  These cores are probably rather oblivious to what
happens around them as they are gravitationally independent entities.
Thus, if such cores survived into the bridge region, we would see the
star formation occurring.

To be ejected into the bridge region, however, the cores have experienced
a sudden deceleration of several 100 $\kms$.  Just as for the rest of the
cloud, the collision necessarily ionizes the dense cores as well.
However, because the probability of a dense core encountering another dense
core is low, the dense core will continue deeper into the incoming cloud,
ending up in a lower density medium than the barycenter of the two clouds.
It may then be able to expand and lose its ``pre-stellar'' nature.
The ionization and subsequent cooling of the clouds may result in some
degree of homogenizing of the clouds, such that time is required
before overdensities can become pre-stellar cores.

Is there any evidence for this in the line ratios?  Only the Taffy
system is strong enough in the CO lines to have any $^{13}$CO or HCN
observations.  \citet{Braine03} found that the $^{13}$CO lines were quite
weak in the Taffy bridge, with line ratios of about 50 and 100 respectively
in the (1--0) and (2--1) transitions respectively.  Taken at face value,
these line ratios indicate that the $^{12}$CO lines are not as optically
thick as they are in galactic clouds or in normal galaxies.  In UGC 12915
the ratios are lower, with ratios of 15 and 19 in the (1--0) and (2--1)
transitions respectively, close to values in normal spirals.  However, in
IR-luminous galaxies with very high SFRs, the $^{12/13}$CO line ratios are
also very high \citep{Casoli92} and the star formation is particularly
efficient.  In the IR-luminous galaxies, the HCN(1--0) line is strong,
sometimes stronger than the $^{13}$CO(1--0) line \citep{Solomon92}.
In the Taffy bridge, we did not detect the HCN(1--0) line but the upper limit
is only slightly below the $^{13}$CO(1--0) line intensity in the bridge.
Thus, the line intensities are compatible with a low opacity in the
bridge region.

\section{Are head-on collisions cosmologically important?}

It is now generally accepted that galaxies were smaller at high
redshift and that local universe galaxies ({\it i.e.} with small
lookback times and hence current objects rather than early universe
objects) are the product of collisions and accretion of smaller entities.
The details of this picture remain quite unclear, however, but these
collisions are mostly tidal rather than the head-on collisions that are
the subject of this work.  We now consider two environments, field
and dense clusters, to briefly explore whether ISM-ISM collisions
could play a role in galaxy evolution for a large number of galaxies.

To keep numbers simple, we will consider spiral galaxies to have a
radius of 10 kpc and take a dense cluster environment to have a density
of 1000 galaxies per Mpc$^3$ at a velocity of 1000 $\kms$.  The field
galaxies are of the same size but have a density of 0.1 galaxies per
Mpc$^3$.  Going through the same mean free path calculation as before,
$$\lambda_{\rm mfp} = 1/(4 A_{gal} n_{gal}) = 1/(4 \pi R^2 n) \approx 800 kpc $$
but at 1000 $\kms$ a galaxy moves 800 kpc in only 800 Myr.
In this extreme case, each dense cluster galaxy undergoes an ISM-ISM
collision every Gyr!  ISM-ISM collisions might then be a very efficient
means of bringing gas out of galaxies and into the intracluster medium,
especially at earlier epochs when clusters were denser and galaxies
more gaseous.  The above numbers are probably overestimates: if the
galaxies touch at a radius of 10kpc, the collision will not bring
gas out of the inner parts -- a more appropriate value of $R$ would be
5 kpc; furthermore, the galaxies should be assumed to have an average
inclination of 45 degrees, reducing the area by $\sqrt{2}$; on the other
hand, since the target galaxies move as well, the effective velocity is
raised by about the same factor.  Nonetheless, each cluster galaxy would
still collide every few Gyrs, such that every cluster galaxy would undergo
a head-on collision at least once and several times for some objects.
Todays rich clusters are spiral-poor but they were certainly much
more spiral-rich in the past, as well as being denser.
The gas ejected from the inner parts of galaxies would naturally be
quite metal-enriched because star formation starts in the inner
parts of spirals.

A similar calculation shows that the situation is entirely different
for field galaxies.  At the much lower density adopted for field spirals,
only about one galaxy out of 1000 suffers this sort of collision in
10 Gyr.  Even allowing for a density increase of $n(z) \approx n_0 (1+z)^3$,
nothing changes out to a redshift of a few at least.
We conclude that while these collisions are likely very important
for galaxy clusters, they do not affect the evolution of field spirals.

\section{Conclusions}

Consistent with the Taffy galaxies, UGC 12914/5, the second such system, UGC 813/6,
also contains large amounts of molecular gas in the bridge region between the
galaxies.  UGC 813 and 816 are considerably smaller and less gas-rich than
UGC 12914 and 12915, and the CO emission from the UGC~813/6 bridge is about
4 times less than from the UGC 12914/5 bridge.  Nonetheless, the CO emission
from the UGC~813/6 bridge is similar to that from the entire Milky Way and the
H$_2$/H{\sc i} mass ratio is roughly unity.

It is shown that for reasonably gas-rich, but not extraordinary, spiral
galaxies a head-on collision will result in many collisions of molecular clouds.
Some earlier calculations had reached the opposite conclusion but they had
supposed very gas-poor disks.  Although the neutral (both atomic and molecular)
clouds are ionized in collisions, the dense gas cools very rapidly allowing it to
recombine and become molecular while still in the colliding disks.  The
new cloud is then left in the bridge because the collision has left it
with an intermediate velocity.

Some H$_2$ may form through collisions of atomic gas clouds just as
some of the originally molecular (or atomic) gas will remain
ionized after collision.  In addition to showing that GMC -- GMC collisions
occur in cases where the nuclei of the galaxies have passed within a
few kpc of each other, several factors lead us to conclude that
GMC -- GMC collisions play a major role in bringing molecular gas into the
bridge regions.  From a theoretical point of view, the study by
\citet{Hollenbach89} indicates that a column density of about $4 \times 10^{21}$
cm$^{-2}$ is necessary to form H$_2$.  This is a large column density for
purely atomic clouds.  Furthermore, if H{\sc i} - H{\sc i} cloud collisions
produced the H$_2$, the molecular gas would be as extended as the H{\sc i}
in the bridge -- this is not the case.  Being much denser than atomic gas
clouds, GMCs will be able to cool much more quickly after collision and
ionization, forming the molecular gas we now observe in the bridge.

The star formation in the bridge material is below that expected
based on the molecular gas mass.  We speculate that the cloud collisions
may have eliminated many of the pre-stellar cores, such that in the few
$10^7$ years since the collision the cores have not had time to reform and
generate the strong stellar luminosity expected.

ISM-ISM collisions may provide an efficient mechanism for driving enriched
gas out of the inner parts of spirals in galaxy clusters.  Given that
clusters were even denser in the past, most cluster galaxies have probably
undergone such a collision at least once.  Only a very small fraction of
field spirals, even at high redshift, would have suffered a head-on collision.

 The future of the bridge
material is as yet unclear.  With a velocity of $\la 300 \kms$ with
respect to one or another of the spirals,
some of the gas is probably destined to fall back onto the parent spirals.
Some however may remain in between the receding galaxies.

\begin{acknowledgements} We would like to thank J.J. Condon for
the H{\sc i} and radio continuum data.  We would also like to
thank Simon Verley of the IAA for taking the R band image used in
figures 1 and 2 and Ra\'ul M\'ujica of the INAOE for the H$\alpha$ image
presented in figure 4.  UL and SL are supported by the Spanish MCyT Grant
AYA 2002-03338 and the Junta de Andaluc\'\i a.  VC would like to acknowledge
the support of JPL contract 960803.
\end{acknowledgements}

\bibliographystyle{apj}
\bibliography{jb}

\end{document}